\documentclass[12pt]{iopart}
\usepackage{graphicx}
\usepackage{cite}

\begin{document}
\title{Identifying structural order in Selenium with Near-Edge Spectroscopy}

\author{J. A. McLeod$^1$, N. Chen$^2$, R. E. Johanson$^3$, G. Belev$^3$, D. Tonchev$^3$, A. Moewes$^1$, and S. O. Kasap$^3$}

\address{$^1$ Department of Physics and Engineering Physics, University of Saskatchewan, Saskatoon, SK, Canada}
\address{$^2$ Canadian Light Source Incorporated, University of Saskatchewan, Saskatoon, SK, Canada}
\address{$^3$ Department of Computer and Electrical Engineering, University of Saskatchewan, Saskatoon, SK, Canada}

\ead{john.mcleod@usask.ca}

\begin{abstract}
We investigate the crystallization of amorphous arsenic-selenium alloys with 0\%, 0.5\%, 2\%, 6\%, 10\%, and 19\% arsenic by atomic concentration using synchrotron X-ray absorption spectroscopy. We identify crystalline order using the extended X-ray absorption fine structure (EXAFS) spectra and correlate this order to changes in features of the X-ray absorption near edge structure (XANES) spectra. We find supporting evidence that the structure of amorphous selenium is composed of disordered helical chains, and is therefore closer to the trigonal crystalline phase than the monoclinic crystalline phase.
\end{abstract}

\section{Introduction}
The photoconducting properties and the low lateral diffusion of photoinduced carriers in amorphous selenium (aSe) have recently led it to find use in commercial direct conversion X-ray imaging devices in medical applications \cite{kasap_2006}. Two open problems with the utility of amorphous selenium as a photoconductor are charge trapping and crystallization. aSe spontaneously crystallizes at room temperature. This process can be delayed by alloying the selenium with between 0.2\% and 0.5\% arsenic by atomic concentration, but such doping increases the charge trapping \cite{kasap_2000}.

The local structure of aSe has been the subject of considerable interest over the years \cite{joannopoulos_1975, hohl_1991, bruning_2001}, and the relation between aSe and the common crystalline phases of monoclinic and tSe (mSe and tSe, respectively) have been discussed \cite{kaplow_1968, laude_1973}. The current consensus is that aSe is mainly composed of disordered helical chains similar to the ordered hexagonal arrangement of helical chains in tSe, with very little contribution from Se$_8$ rings (which form the basis for mSe)\cite{baganich_1991, jovari_2003}.

Herein we attempt to induce crystallization in arsenic-selenium alloys and use X-ray absorption near edge structure (XANES) and extended X-ray absorption fine structure (EXAFS) spectra recorded at the selenium \textit{K}-edge to probe the local structure. We find evidence supporting the belief that aSe is composed of helical chains rather then Se$_8$ rings, and we identify XANES features which act as fingerprints identifying the degree of crystallization in the material.

\section{Experimental Procedure}
The samples were prepared by vapour deposition of pure (99.999\%) selenium and arsenic-selenium alloys on a room temperature polycarbonate substrate \cite{kasap_2003}. The different arsenic concentrations in the arsenic-selenium alloys were 0.5\%, 2\%, 6\%, 10\% and 19\% by atomic composition. The vapour deposition process formed homogenous films between 24 $\mu$m and 35 $\mu$m thick. This range of thickness was chosen to give an absorption step of roughly 2 at the the selenium $K$-edge. 

As reference standards, tSe crystals were grown from vapour produced due to the sublimation of pure selenium kept at a temperature of 200$^\circ$C in a closed glass vessel for 7-10 days. The crystals grown by this method have needle like shape that is typical for tSe \cite{brown_1914}. mSe crystals were grown by a  saturated solution of selenium in methylene iodide (CH$_2$I$_2$) \cite{grunwald_1973}. The tSe crystals were ground into a fine powder (less then 50 $\mu$m grain size). The mSe crystals were essentially grown in powder form, and were not further modified since mechanical stress is known to induce conversion to the trigonal phase \cite{guo_1998}. These powders were homogeneously mixed with boron nitride. The ratio of selenium to boron nitride was chosen to produce a similar absorption step to that of the arsenic-selenium alloy films.

Some of the film samples were annealed in a convection oven to induce crystallization. The films produced samples with 6\% or less arsenic were annealed at 100$^\circ$C for 24 hours. The annealing temperature was substantially higher than the highest glass transition temperature for these materials (about 80$^\circ$C for the Se:6\%As alloy). The films  with 10\% and 19\% arsenic were annealed for about 100 hours at 110$^\circ$C (the maximum temperature of the furnace). This temperature was only slightly higher than the highest glass transition temperature (about 106$^\circ$C for Se:19\%As). Finally, aSe films were measured as prepared, and then remeasured immediately after annealing at 60$^\circ$C, 80$^\circ$C, and finally 100$^\circ$C for three hours each.

The X-ray measurements were performed at the Hard X-ray Microprobe Analysis (HXMA) beamline of the Canadian Light Source (CLS) \cite{hxma_ref}. The measurements were performed with a Si(220) crystal monochromator and an Rh-coated harmonic rejection mirror. The measurements were performed in transmission mode. A liquid helium cryostat was used for the low temperature measurements, the temperature was stabilized between 30 K and 45 K. Since no difference was observed between the room temperature and low temperature XANES measurements, the measurements of the progressively annealed pure selenium were conducted at room temperature. A standard pure selenium film was measured concurrently with all spectra as a reference for calibration. The absorption edge of this reference was calibrated to 12658 eV using the zero-crossing in the second derivative nearest the edge.
\begin{figure}[ht]
\begin{minipage}{16pc}
\includegraphics[width=16pc]{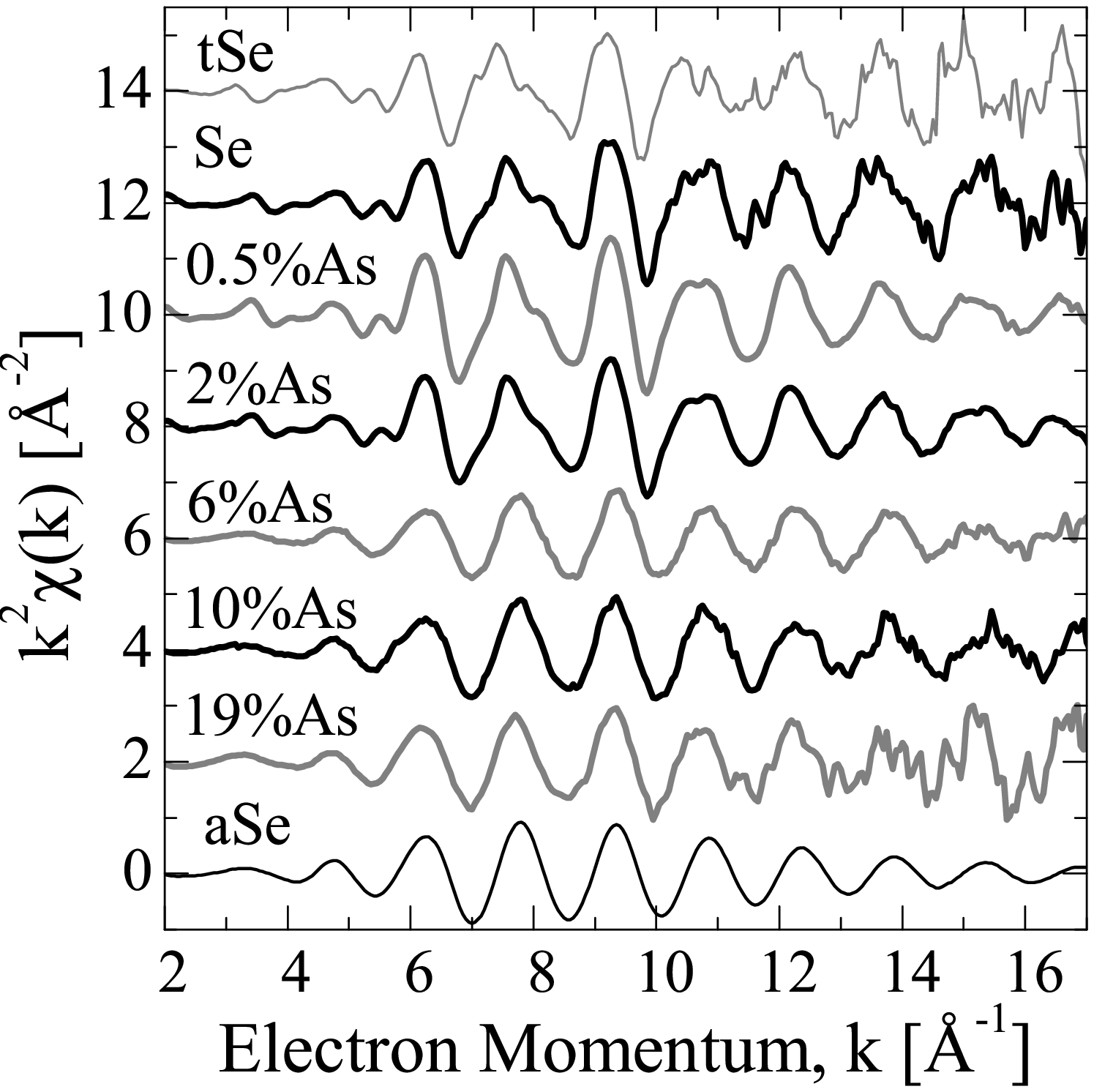}
\end{minipage}\hspace{2pc}
\begin{minipage}{16pc}
\includegraphics[width=16pc]{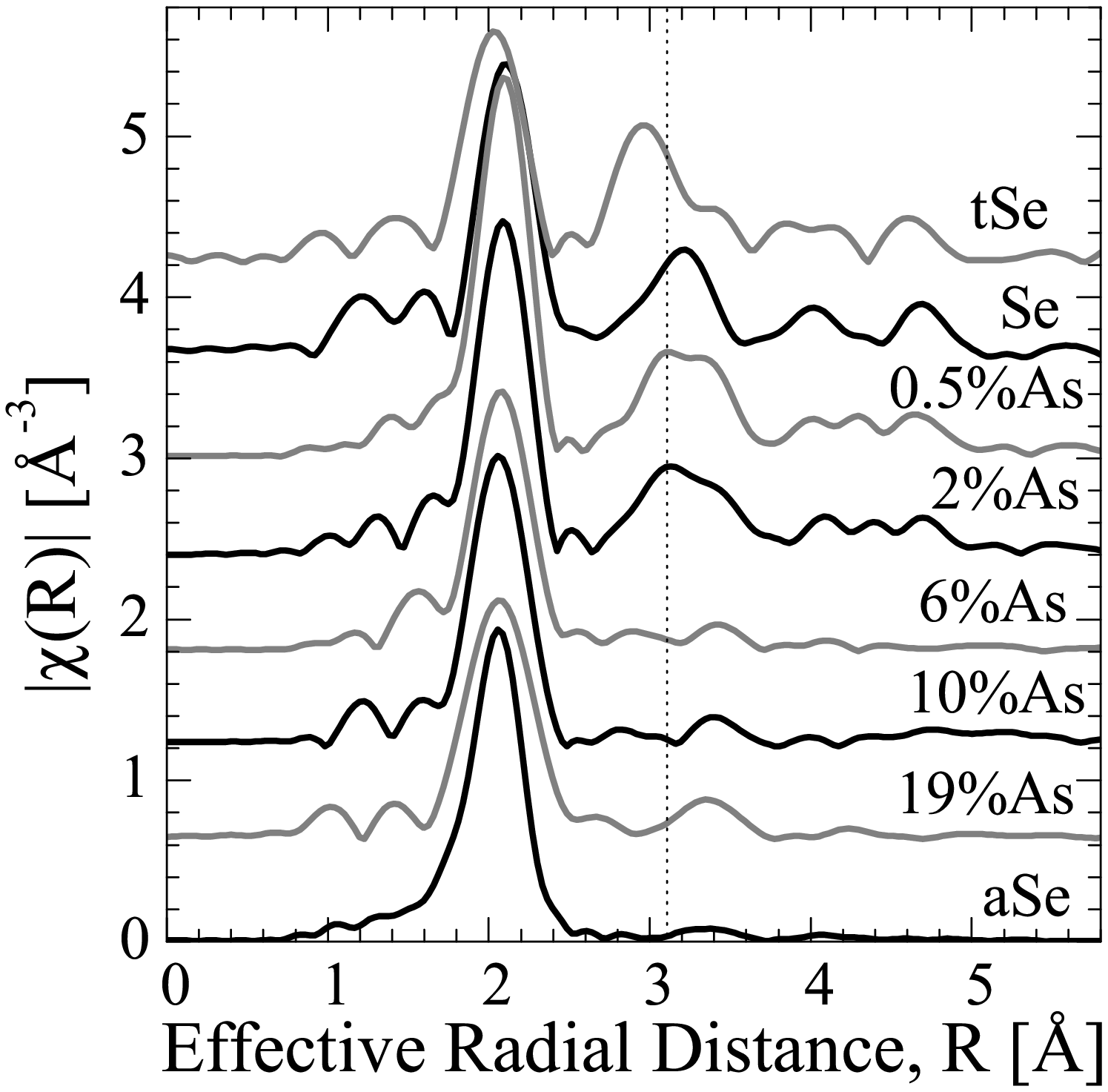}
\end{minipage}
\caption{\label{fig:se_exafs}The EXAFS oscillations, $k^2\chi(k)$ and the magnitude of the Fourier transform, $\vert \chi(R)\vert$ for the low-temperature selenium \textit{K}-edge measurements of annealed samples and unannealed tSe, and aSe. Note the nearest neighbour appears at roughly 2.1 \AA, the actual bondlength is about 2.34 \AA \cite{majid_1998}.}
\end{figure}

\section{Results and Discussion}
The EXAFS of the low temperature selenium \textit{K}-edge measurements reveal some long range order in some of the annealed samples. Figure \ref{fig:se_exafs} shows that in samples with an arsenic concentration of less than 6\% the second nearest neighbour at roughly 3 \AA is readily apparent. This medium range order in the EXAFS indicates that the samples with less than 6\% arsenic are polycrystalline. There is a weak feature at about 3.3 \AA in samples with arsenic concentration greater than 6\%, but this feature is also present in aSe. Annealing does increase the intensity of this feature, but it is still much weaker than the medium range features in samples with less than 6\% arsenic. The shape of the EXAFS $\vert \chi(R)\vert$ is consistent with literature reports on the EXAFS of tSe \cite{kolobov_1996}, and this suggests that the polycrystalline arsenic-selenium alloys have some degree of helical chain structure consistent with the structure of tSe \cite{jovari_2003}. To contrast, the EXAFS of mSe has the second nearest neighbour at about 3.6 eV \cite{endo_1993}, a feature which is not present in any of the annealed samples here.

\begin{figure}[ht]
\begin{minipage}{16pc}
\includegraphics[width=16pc]{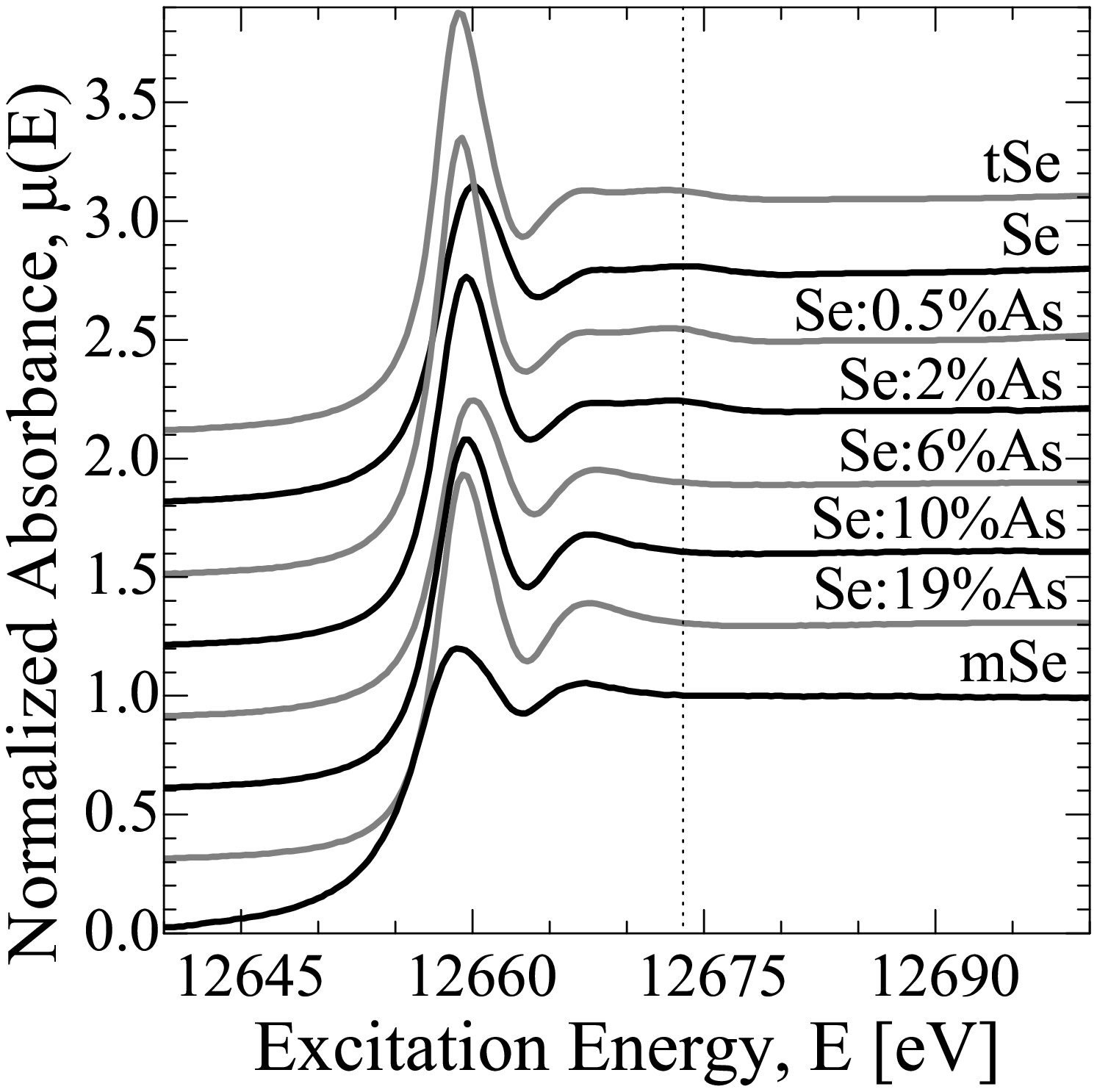}
\caption{\label{fig:se_xanes_1}The selenium \textit{K}-edge XANES measurements for low-temperature arsenic-selenium alloys annealed in a convection oven. Note the spectral difference between low-arsenic ``polycrystalline'' alloys and the high-arsenic amorphous alloys.}
\end{minipage}\hspace{2pc}%
\begin{minipage}{16pc}
\includegraphics[width=16pc]{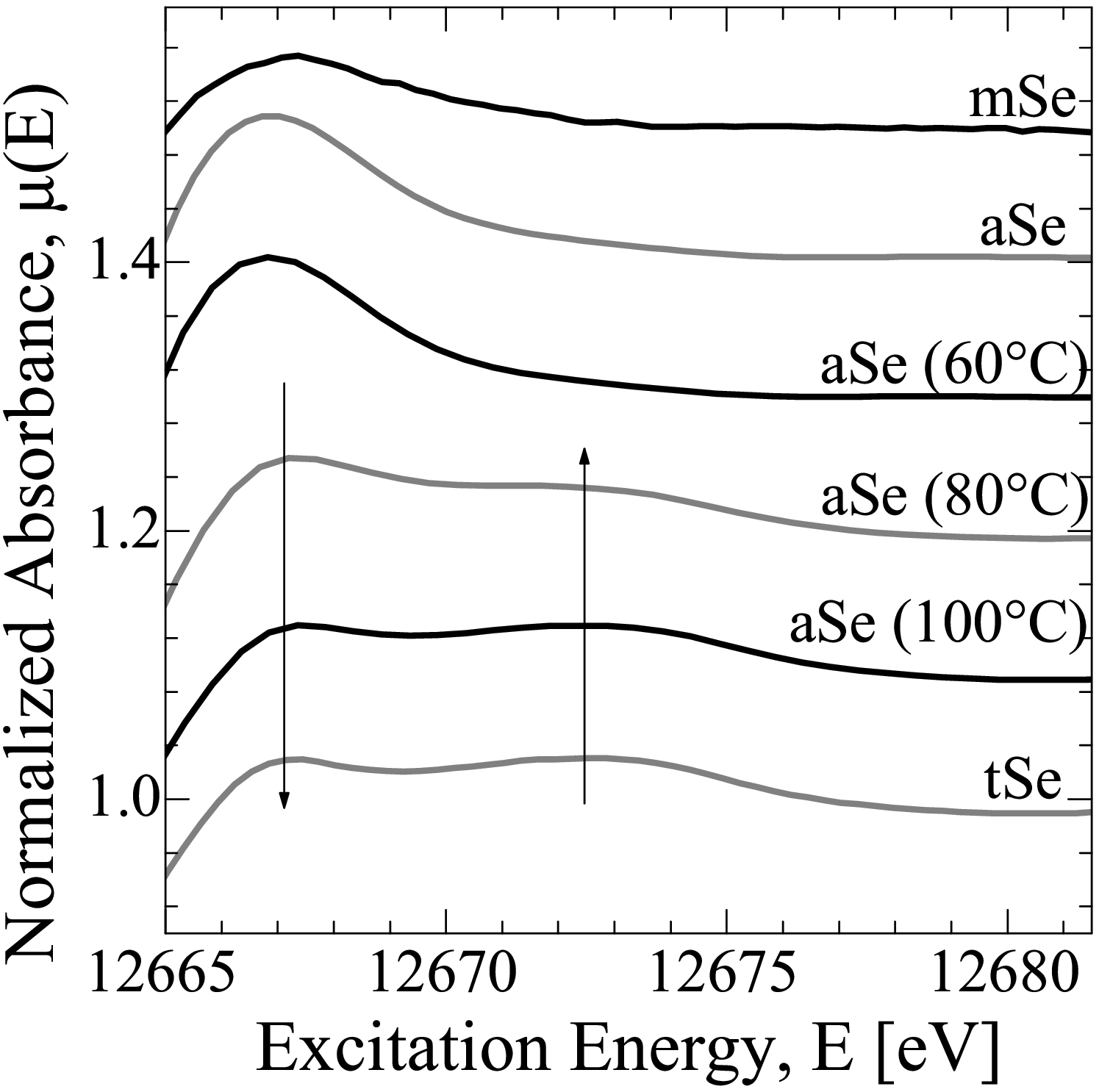}
\caption{\label{fig:se_xanes_2}Some selenium \textit{K}-edge XANES measurements for room-temperature materials. The range of the spectra highlights the growth of the post-edge feature at $\sim$12673 eV.}
\end{minipage} 
\end{figure}

The XANES of the low-temperature selenium \textit{K}-edge measurements shows the prominent white line (at roughly 12660 eV) and secondary post-edge feature (at roughly 12668 eV) that are common to both trigonal, monoclinic, and amorphous pure selenium \cite{buchanan_1995}.  Figure \ref{fig:se_xanes_1} shows that there is a minor tertiary post-edge feature at roughly 12673 eV that occurs in the annealed samples with less than 6\% arsenic that does not occur in the other annealed samples. Since the EXAFS spectra reveal that the samples with less than 6\% arsenic are polycrystalline, this suggests that this post-edge XANES feature is connected to polycrystalline order. 

The post-edge shape of the mSe XANES resembles that of the amorphous arsenic-selenium alloys, while the the post-edge shape of the tSe XANES resembles that of the polycrystalline arsenic-selenium alloys. The shape of the post-edge XANES has been previously proposed used as a fingerprint to identify the phase of biologically produced selenium in Reference  \cite{buchanan_1995}, but this study did not recognize that aSe shares the same XANES features as mSe. 

Figure \ref{fig:se_xanes_2} displays the development of this post-edge XANES feature. As pure selenium is progressively annealed at higher temperatures, the post-edge XANES feature gradually changes from that of aSe to that of tSe. Since recent studies on the structure of aSe indicate that it does not have a significant amount of Se$_8$ rings \cite{jovari_2003}, the similarity between the post-edge XANES features in amorphous and mSe should not be interpreted as an indication of local ``monoclinic-like'' structure. On the contrary, since it is generally accepted that aSe is composed of disordered helical chains \cite{baganich_1991} and since the similarity in post-edge XANES features between annealed arsenic-selenium alloys and tSe is concurrant with a similarity in the EXAFS, we propose that the post-edge XANES features can be an indication of local ``trigonal-like'' structure.

To summarize, we have measured the EXAFS and XANES spectra of several arsenic-selenium alloys after annealing in a convection oven. We have identified that materials with a low concentration of arsenic change to a polycrystalline phase composed of helical chains, similar to tSe. The threshold for this polycrystalline phase change is between 2\% and 6\% arsenic. We have also identified a post-edge XANES feature that we suggest may be used as a quick fingerprint to identify the degree of ``trigonal-like'' polycrystallinity in arsenic-selenium alloys.

\ack{
We acknowledge the support of the Natural Sciences and Engineering Research Council of Canada (NSERC) and the Canada Research Chair program. The research described in this paper was performed at the Canadian Light Source, which is supported by NSERC, NRC, CIHR, and the University of Saskatchewan.}

\bibliographystyle{iopart-num}
\bibliography{aSe_XAFS14}

\providecommand{\newblock}{}
\begin{thebibliography}{10}
\expandafter\ifx\csname url\endcsname\relax
  \def\url#1{{\tt #1}}\fi
\expandafter\ifx\csname urlprefix\endcsname\relax\def\urlprefix{URL }\fi
\providecommand{\eprint}[2][]{\url{#2}}

\bibitem{kasap_2006}
Kasap S~O, Kabir M~Z and Rowlands J~A 2006 {\em Curr. Appl. Phys.\/} {\bf 6}
  288

\bibitem{kasap_2000}
Kasap S~O 2000 {\em J. Phys. $\mathrm{D}$: Appl. Phys.\/} {\bf 33} 2853

\bibitem{joannopoulos_1975}
Joannopoulos J~D, Schl\"{u}ter M and Cohen M~L 1975 {\em Phys. Rev.\/} B {\bf
  11}(6) 2186

\bibitem{hohl_1991}
Hohl D and Jones R~O 1991 {\em Phys. Rev.\/} B {\bf 43}(5) 3856

\bibitem{bruning_2001}
Br\"{u}ning R, Irving E and LeBlanc G 2001 {\em J. Appl. Phys.\/} {\bf 89}(6)
  3215

\bibitem{kaplow_1968}
Kaplow R, Rowe T~A and Averbach B~L 1968 {\em Phys. Rev.\/} {\bf 168}(3) 1068

\bibitem{laude_1973}
Laude L~D, Kramer B and Maschke K 1973 {\em Phys. Rev.\/} B {\bf 8}(12) 5794

\bibitem{baganich_1991}
Baganich A~A, Mikla V~I, Semak D~G {\em et~al.\/} 1991 {\em phys. stat. sol.
  (b)\/} {\bf 166} 297

\bibitem{jovari_2003}
J\'{o}v\'{a}ri P, Delaplane R~G and Pusztai L 2003 {\em Phys. Rev.\/} B {\bf
  67} 172201

\bibitem{kasap_2003}
Kasap S~O, Koughia K~V, Fogal B {\em et~al.\/} 2003 {\em Semiconductors\/} {\bf
  37}(7) 789

\bibitem{brown_1914}
Brown F~C 1914 {\em Phys. Rev.\/} {\bf 4}(2) 85

\bibitem{grunwald_1973}
Grunwald H~P 1973 {\em Mater. Res. Bull.\/} {\bf 7}(10) 1093

\bibitem{guo_1998}
Guo F~Q and Lu K 1998 {\em Phil. Mag. Lett.\/} {\bf 77}(4) 181

\bibitem{hxma_ref}
Jiang D~T, Chen N and Sheng W 2007 {\em AIP Conference Proceedings\/} vol 879 p
  800

\bibitem{majid_1998}
Majid M, B\'{e}nazeth S, Souleau C and Purans J 1998 {\em Phys. Rev.\/} B {\bf
  58}(10) 6104

\bibitem{kolobov_1996}
Kolobov A~V, Oyanagi H, Tanaka K and Tanaka K 1996 {\em J. Non-Crys. Solids\/}
  {\bf 198-200} 709

\bibitem{endo_1993}
Endo H, Maruyama K, Tsuzuki T and Yao M 1993 {\em Jpn. J. Appl. Phys.\/} {\bf
  32} 773 suppl. 32-2

\bibitem{buchanan_1995}
Buchanan B~B, Bucher J~J, Carlson D~E {\em et~al.\/} 1995 {\em Inorg. Chem.\/}
  {\bf 34} 1617

\end{thebibliography}
\end{document}